\documentclass[prb,showpacs,twocolumn]{revtex4}
\usepackage{epsfig}
\begin{document}
\bibliographystyle{revtex}
\title{Voltage induced conformational changes and current control\\
 in charge transfer through molecules}
\author{Lars Kecke}
\affiliation{Institut f\"ur theoretische Physik, Universit\"at Ulm,
Albert-Einstein-Allee 11, D-89069 Ulm, Germany}
\author{Joachim Ankerhold}
\affiliation{Institut f\"ur theoretische Physik, Universit\"at Ulm,
Albert-Einstein-Allee 11, D-89069 Ulm, Germany}

\date{\today}

\begin{abstract}
Transport through molecular contacts with a sluggish intramolecular vibrational
mode strongly coupled to excess charges is studied far from equilibrium.
A Born-Oppenheimer approximation in steady state reveals voltage dependent
energy surfaces which cause abrupt conformational changes of the molecular
backbone. These are directly related to transitions between current plateaus
which are relatively robust against thermal fluctuations. In a regime accessible in experiments
this allows to operate a molecular junction as a current switch or as a molecular machine in form
of a valve controlled by time dependent bias and gate voltages.

\end{abstract}
\maketitle

\section{Introduction}
In recent years impressive progress has been achieved in fabricating,
functionalizing and controlling molecular junctions  for charge transfer
[\onlinecite{reviews,cuniberti,scheer}] and as nano-machines
[\onlinecite{browne}].
This includes advanced techniques to link molecular structures to metallic
electrodes and gates, to design molecular rectifiers or to induce rotary
molecular motion. Particularly fascinating is the interplay of charge transfer
and conformational dynamics.

Theoretically, density functional calculations (DFT) in combination with
non-equilibrium Greens function techniques as well as model based descriptions
in terms of master equations have elucidated fundamental aspects of single
molecule contacts.
A particular challenge is to capture electromechanical properties of these
devices where injected charges drive internal molecular degrees of freedom
such as vibrational or rotational modes far from equilibrium which in turn
back-act on the transport channels. A well-established procedure is to
determine energy surfaces for these modes from the conformation of the
molecular backbone in or at least very close to equilibrium. Since couplings
between molecules and electrodes are typically weak and relevant
intra-molecular excitations are fast compared to displacements of the
skeleton of the entire structure, this approach has been successfully
applied to explain a variety of experimental observations
[\onlinecite{scheer}].
What happens, however, far from equilibrium is much less understood.

To address this issue we consider in this paper a minimal system consisting of
a conjugated organic molecule with two sub-units the relative orientation of
which determines its transfer properties when contacted to electrodes. A
typical realization is e.g.\ a biphenyl compound with the dihedral angle
between the two benzene rings acting as conformational degree of freedom. This set-up
has received considerable attention in recent experiments, see
e.g.\ [\onlinecite{dadosh,venka,mis1,mis2}]. So far, however, only situations close
to equilibrium (low voltage regime) and for very weak electrode-molecule
couplings have been addressed. Theoretical work has revealed the role of
vibrational excitations [\onlinecite{thoss}] and dynamical symmetry breaking
[\onlinecite{grifoni}] within Master equation approaches. For compounds with
fixed torsion mode DFT calculations apply to show microscopic details of conductance properties
[\onlinecite{xue,xia,wang}].

Here we focus on the complementary regime of strong non-equilibrium with a
rotational mode that is not frozen, but may approach a voltage dependent
optimal configuration in steady state. This scenario requires
electrode-molecule couplings large compared to vibrational but small compared
to electronic excitations on the molecule (details are given below). Then, a
Born-Oppenheimer approximation (BO) applies which provides voltage
dependent energy surfaces. Notably, stable molecular conformations can be
controlled by bias and gate voltage which are directly related to a robust
quantization of current transfer. This opens the possibility for realizations
of molecular machines [\onlinecite{browne,selden}] as e.g. switches and valves.

 For charge transfer through molecular contacts BOs in steady state have been discussed previously in several contexts. In [\onlinecite{nitzan}] a type of self-consistent treatment for a contact with a single electronic level coupled to single harmonic intramolecular degree of freedom  revealed the appearance of hysteresis and negative differential conductance. Greens functions methods have been used in conjunction with a BO
to extract effective potential surfaces for nearly harmonic vibronic modes in [\onlinecite{greens}].
 However, both treatments exploit that the driven dynamics of an {\em harmonic} oscillator is analytically known.
 Instead, in the sequel a methodology for substantial anharmonicity of intramolecular modes and/or Coulomb interaction between excess charges on several molecular electronic levels is studied.

\section{Model}
\begin{figure}
\epsfig{file=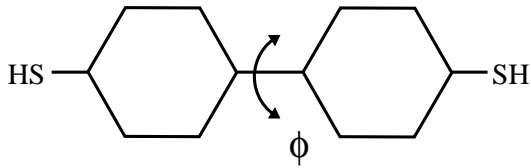, width=7cm}
\caption{Sketch of the biphenyl molecule showing the rotational angle $\phi$.
Connection to the electrodes is established via the thiole groups.}\label{sketch}
\end{figure}
The biphenyl molecule (Fig.~\ref{sketch}) consists of two benzene rings which
can be twisted by an angle $\phi$. Each of the rings may carry one excess electron
and the tunnel coupling between left ring (L) and  right ring (R) depends on
this torsion angle as the dominant intramolecular mode with respect to
transport properties. The total molecule Hamiltonian then reads [\onlinecite{thoss}]
\begin{eqnarray}
\label{molecule}
H_0&=&-\frac{1}{2 I}\frac{\partial^2}{\partial\phi^2}+V(\phi)
+T(\phi)(d_L^\dagger d_R+d_R^\dagger d_L)
\nonumber\\
&&+E_c\, (n_L+n_R)+U_0\, n_L n_R+\frac{V_S}{2}(n_L-n_R)\, ,
\end{eqnarray}
where the first line describes the dynamics of the torsion angle and tunneling
between the rings, i.e.,
\begin{equation}
 V(\phi)=V_0\cos^2(2\phi)\, , \ T(\phi)=T_0\cos\phi\ ,
\end{equation}
and the second line captures the charging energy $E_c$ for individual excess
charges on the molecule and the Coulomb repulsion $U_0$ for double occupancy,
respectively.
Following microscopic calculations
[\onlinecite{xue}] we also incorporate a Stark energy
shift $V_S$ of the molecular sites due to an external electric field. It turns out
that this Stark shift has substantial impact on the molecular conformation and
transport properties.
Further we have the local electron densities $n_\alpha=d_\alpha^\dagger d_\alpha,
\alpha=L, R $, where operators $d_L^\dagger$  ($d_R^\dagger$) create excess
electrons on the left (right) subunit.
Crucial for the charge transfer across this structure is the separation of
energy scales with a small height $V_0=0.05{\rm eV}$ for the rotational barrier
of the neutral molecule, a large hopping element $T_0=0.5{\rm eV}$, and a large
inertial moment $I=20000 {\rm eV}^{-1}$ [\onlinecite{ssp1,ssp2}]. As it is
typical for conjugated organic molecules, orbitals are thus delocalized
throughout the molecular backbone.  Accordingly, it is appropriate to introduce
the molecule's charge energy eigenstates as linear combinations of the left
and right localized states:
$d_+^\dagger=\cos\theta d^\dagger_L+\sin\theta d^\dagger_R$
and $d_-^\dagger=-\sin\theta d^\dagger_L+\cos\theta d^\dagger_R$.
The mixing angle $\theta\in[-\pi/2,\pi/2]$ depends on the torsional degree of freedom
\begin{equation}
\theta(\phi)=\arctan\left(\frac{T(\phi)}{\lambda(\phi)+\frac{V_S}{2}}\right)\,,
\label{thetaphi}
\end{equation}
with the eigenenergies of the one-electron Hamiltonian
$\lambda(\phi)=\sqrt{T(\phi)^2+V_S^2/4}$.
As one can see, for vanishing Stark shift, we have $\theta=\pm\pi/4$
(the sign of $\theta$ jumps at $\phi=\pi$ where the hopping matrix elemnt crosses
zero) and the energy eigenstates become the symmetric and antisymmetric combinations
of localized states, which couple symmetrically (except for a sign) to the right
and left leads. For large Stark shift we instead get $\theta=0$ and the
eigenstates become the localized ones that only couple to a single lead each.

In eigenstate representation the molecule Hamiltonian is
\begin{eqnarray}
H_0&=&-\frac{1}{2 I}\frac{\partial^2}{\partial\phi^2}+E_0(\phi)\,
|0\rangle\langle 0|+E_+(\phi)\, |+\rangle\langle +|\nonumber\\
&&+E_-(\phi)\, |-\rangle\langle -|+E_D(\phi)\, |D\rangle\langle D|\, .
\end{eqnarray}
Here, $|0\rangle$ denotes the neutral molecule, $|\pm\rangle$ the two
one-particle eigenstates, and $|D\rangle$ is the doubly occupied state.
The torsional dependent energies of these states are given by:
$E_0(\phi)=V(\phi)$, $E_\pm(\phi)=V(\phi)\pm \lambda(\phi)+E_c$, and
$E_D(\phi)=V(\phi)+2E_c+U_0$.

\begin{figure}
\epsfig{file=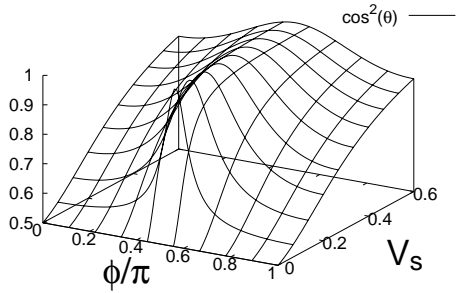, ,angle=270,width=7cm}
\caption{Frank-Condon-factor $\cos^2\theta(\phi)$ as a function of $\phi$ and
$V_s$. As described in the text, a finite electric field breaks the symmetry
of the one-electron eigenstates ($\theta(\phi)=\pi/4$) and induces a preference
for angles $\phi\approx\pi/2$ in the coupled system.}\label{coupfig}
\end{figure}

The molecular junction is now modeled by the Hamiltonian $H=H_0+H_I+H_L+H_R$
with $H_\alpha = \sum_k (\epsilon_{k\alpha} -\mu_\alpha) c_{k\alpha}^\dagger
c_{k\alpha}, \alpha\in\{L,R\}$ describing left (L) and right (R) lead,
respectively, as reservoirs of non-interacting quasi-particles with creation
(annihilation) operators $c_{k\alpha}^\dagger$ ($c_{k\alpha}$).  The chemical
potentials $\mu_{L/R}=V_g\pm V_b/2$ are fixed by the bias voltage $V_b$ and the
gate voltage $V_g$. The Stark-shift varies with the bias according to
$V_S=\kappa\, V_b$ with a parameter $\kappa<1$.  Coupling of the left (right)
lead to the left (right) ring is described by
\begin{eqnarray}
\label{coupling}
H_I&=&\gamma_L d_L^\dagger \psi_{L}+\gamma_R d_R^\dagger \psi_{R}+h.c.\nonumber\\
&=&\gamma_L(\cos\theta d_+^\dagger-\sin\theta d_-^\dagger)\psi_{L}
\nonumber\\
&&+\gamma_R(\sin\theta d_+^\dagger+\cos\theta d_-^\dagger)
\psi_{R}+h.c.\,
\end{eqnarray}
with $\psi_\alpha=\sum_k c_{\alpha, k}$.
Charge transfer into/out of the molecule is determined by rates
$\Gamma_\alpha=D_\alpha \gamma_\alpha^2/2\hbar$ where electrodes are taken in the
wide band limit with a density of states $D_\alpha$. Apparently, the coupling of the molecular states depends on the mixing angle $\theta$ which itself depends sensitively on the Stark-field according to (\ref{thetaphi}). For instance,
the coupling of the $|+\rangle$ state to the left (right) lead is given by
$2\Gamma_L\cos^2(\theta)\quad [2\Gamma_R\sin^2(\theta)]$. As shown in
Fig.~\ref{coupfig}, for $\Gamma_L=\Gamma_R\equiv \Gamma$ these couplings
are identical only for vanishing electric field, i.e.\ $V_S=0$ and $\theta(\phi)=\pi/4$. For finite values of $V_S$ this symmetry is immediately broken, e.g.\  around $\phi=\pi/2 -\delta$ with small deviations $\delta$, one has
$\theta(\pi/2-\delta)\approx (T_0/V_S)\, \delta$ so that $2\Gamma \cos(\theta)^2\approx 2\Gamma\gg
2\Gamma\sin^2 \theta\approx 2\Gamma (T_0 \delta/V_S)^2$. 

In the remainder we are particularly interested in the steady state current
$\langle I_\alpha\rangle = e {\rm Tr}\{W_0\dot N_\alpha({t\to \infty})\}
=e {\rm Tr}\{N_\alpha\dot{W}({t\to \infty})\}$ where
$N_\alpha=\sum_k c_{\alpha, k}^\dagger c_{\alpha, k}$ is the number operator in
lead $\alpha$ and $W(t)$ is the density operator  of the junction.
With electrodes residing in thermal equilibrium the reduced density operator
$\rho(t)= {\rm Tr}_{\rm Leads}\{W(t)\}$ provides the relevant transport
information.  A standard procedure  is to derive from the Liouville-von Neuman
equation $\dot{\rho}(t)=(-i/\hbar){\rm Tr}_{\rm Leads}[H,W(t)]$ a master
equation by treating $H_I$ as a perturbation (weak coupling).
This way, upon employing the usual Born-Markov approximation one arrives at
$\dot{\rho}(t)=(-i/\hbar)[H_0,\rho(t)]+\hat{R}\rho(t)$ with the Redfield
tensor[\onlinecite{weiss,mitra}]
\begin{equation}
\hat{R}\rho(t)=-\frac{1}{\hbar^2}\int_0^\infty{\rm d}\tau{\rm Tr}_{\rm Leads}
[H_I,[H_I(-\tau),W(t)]]\,.
\label{redfield}
\end{equation}
Positively definite steady-state solutions are only obtained if a secular
approximation is applied to (\ref{redfield}) so that couplings between
populations (diagonal elements) and coherences (off-diagonal elements) of
{\em non-degenerate} states are dropped [\onlinecite{semigroups}]. The
corresponding Redfield tensor $R$ determines the steady state density due to
$R\rho_{\rm st}=0$.  Further, from(\ref{redfield}) one has
\begin{equation}
\langle I_\alpha\rangle=-\frac{1}{\hbar}\lim_{t\to \infty}\int_0^\infty{\rm d}\tau
{\rm Tr}\{W(t)[I_\alpha,H_I(-\tau)]\}\label{ieq}
\end{equation}
with  the current operator $I_\alpha=(i e/\hbar)[N_\alpha,H_I]=
(i e\gamma_\alpha/\hbar)
\sum_k( d_\alpha c_{k\alpha}^\dagger- d_\alpha c_{k\alpha}^\dagger)$. Of
course, in steady state $I_R=-I_L$.
\section{Steady state observables}
\begin{figure}
\epsfig{file=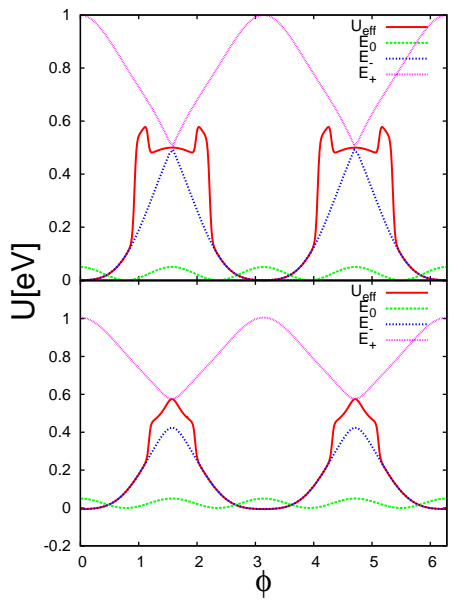, angle=270, width=6.5cm}
\caption{Energy surfaces of the bare molecular states $|0\rangle, |\pm\rangle$
and the BOS $V_{\rm eff}$ vs.\ the torsional angle $\phi$ for $V_b=0.5V$,
$V_g=0.5V$. In upper part bare molecular eigenstates are used, in the lower one
a Stark shift between the left and right site of $V_S=0.3V_b$ is assumed.}\label{eeff}
\end{figure}

For low voltages ($|V_b|, |V_g|<V_0$) and very
weak couplings, transport properties are restricted to the low energy sector.
Thus a representation of $\rho(t)$ in the basis $\{|0, m\rangle, |-, n\rangle\}$
is appropriate, with vibrational modes $\{|m\rangle, |n\rangle\}$ probing only
the vicinity of $\phi=0$ on surfaces $E_0(\phi), E_-(\phi)$, respectively
[\onlinecite{thoss,grifoni}] (see Fig.~\ref{eeff}).

In regime of higher voltages, however, this scheme
poses severe difficulties. On surfaces $E_0, E_\pm$ the density of torsional
eigenstates (typical level spacings $\sqrt{V_0/I}\approx meV$) strongly
increases such that they should more accurately  be described in terms of
coherent states. This necessitates the inclusion of coherences in
(\ref{redfield}) also for {\em non-degenerate} eigenstates which typically
leads to negative populations and thus to a breakdown of the perturbative
treatment. A separation of time scales between the sluggish motion of the
dihedral angle and a faster charge transfer through the contact suggests an
alternative approach though which avoids this deficiency. In the spirit of a
BO one first calculates steady-state solutions
$\rho_{\rm st}(\phi)$ for the electronic system at any {\em fixed} value of the
rotational angle and then uses this solution to extract an effective
steady-state potential for the torsional mode. Within the basis
$\{|0\rangle,|\pm\rangle, |D\rangle\}$ and for given $\phi$  corresponding
energies  are {\em not} degenerate [apart for $V_S=0$ at
$\phi=\frac{\pi}{2} \ {\rm mod}\, \pi$ (Fig.~\ref{eeff})].
A secular approximation applied for fixed $\phi$ to (\ref{redfield}) leads
thus in steady state to
$R (\rho_{00},\rho_{++},\rho_{--},\rho_{DD})_{\rm st}^{t}=0$ with
\begin{equation}
R=\left(\begin{array}{cccc}
\sigma_{00}&-\Sigma^{\rm out}_{+0}&-\Sigma^{\rm out}_{-0}&0\\
-\Sigma^{\rm in}_{+0}&\sigma_{++}&0&-\Sigma^{\rm out}_{D+}\\
-\Sigma^{\rm in}_{-0}&0&\sigma_{--}&-\Sigma^{\rm out}_{D-}\\
0&-\Sigma^{\rm in}_{D+}&-\Sigma^{\rm in}_{D-}&\sigma_{DD}
\end{array}
\right)\, .
\end{equation}
Here, transition rates read $\sigma_{00}=\Sigma^{\rm in}_{+0}
+\Sigma^{\rm in}_{-0}$, $\sigma_{\pm \pm}=\Sigma^{\rm out}_{\pm 0}
+\Sigma^{\rm in}_{D\pm}$, and  $\sigma_{DD}=\Sigma^{\rm out}_{D+}
+\Sigma^{\rm out}_{D-}$, where incoming self-energies are given  by
\begin{equation}
\label{rates}
\Sigma^{\rm in}_{ab}=\sum_\alpha
\Sigma^{\rm in}_{ab,\alpha}=\hbar\sum_\alpha
\Gamma_{ab,\alpha} f_\alpha[E_a(\phi)-E_b(\phi)]
\end{equation}
and outgoing ones by $\Sigma^{\rm out}_{ab}=\hbar\sum_\alpha
\Gamma_{ab,\alpha}-\Sigma^{\rm in}_{ab,\alpha}$.
Further, we have introduced Fermi distributions
$f_\alpha(E)=1/\{1+\exp[\beta(E-\mu_\alpha)]\}$ and
coupling rates:
$\Gamma_{ab,\alpha}=2\,\Gamma_\alpha|\langle a|d_\alpha+d^\dagger_\alpha|b\rangle|^2$.
As noted above, these coupling rates are proportional to $\sin^2\theta(\phi)$
and $\cos^2\theta(\phi)$, respectively, and reduce to $\Gamma_\alpha$ in
absence of a Stark-shift ($V_S=0: \theta=\pi/4$).

\begin{figure}
\epsfig{file=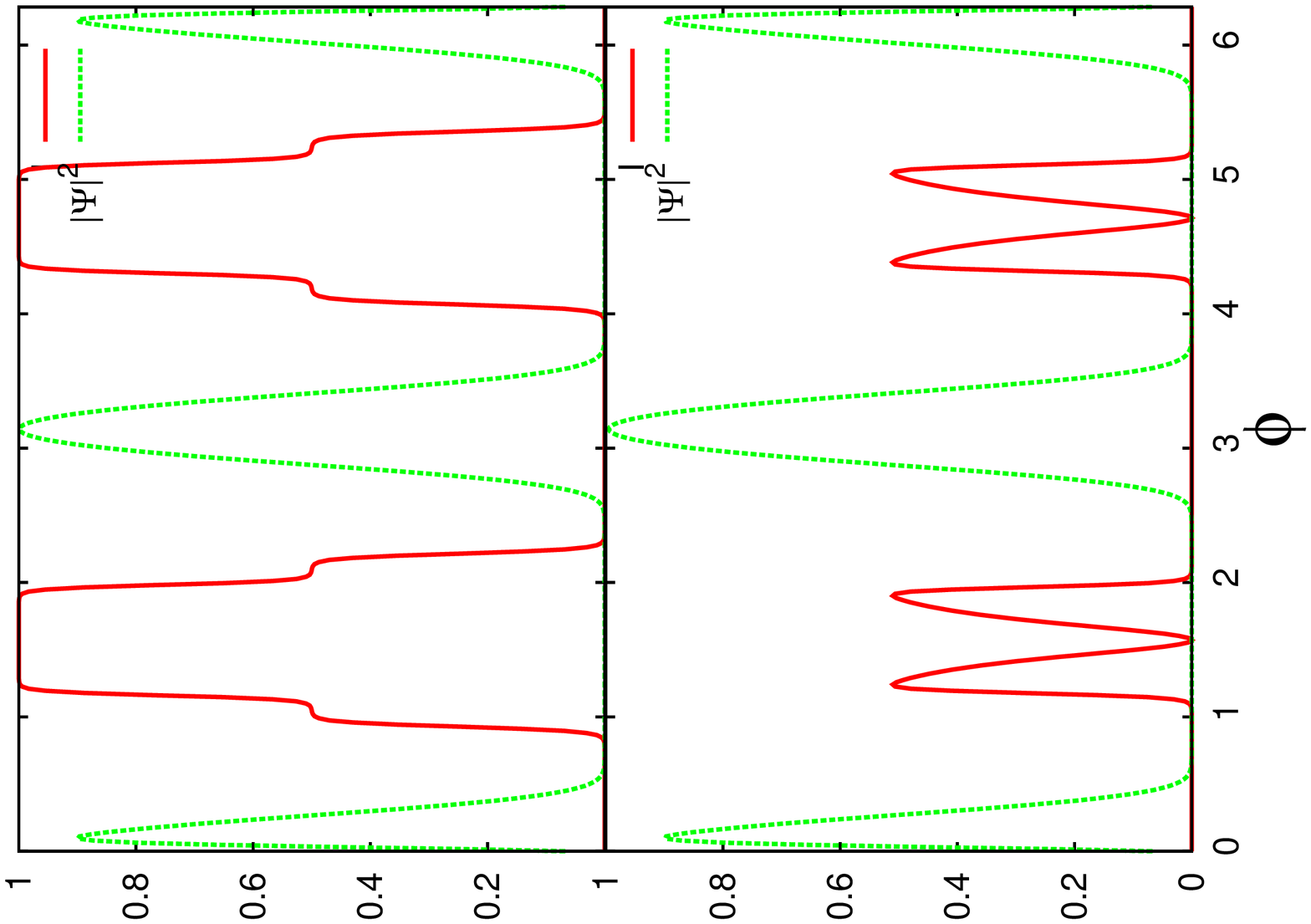, angle=270, width=6.5cm}
\caption{Angle dependent current and probability density for the $T=80K$ rotational
state. Parameters are the same as in Fig.~\ref{eeff}}\label{iphifig}
\end{figure}

The angle dependent steady state density determines all relevant quantities
such as the Born-Oppenheimer surface (BOS) for the torsional mode
\begin{equation}
V_{\rm eff}(\phi)= \sum_{j\in\{0,+,-,D\}}E_j(\phi)\rho_{jj}(\phi)\,
\label{veff}
\end{equation}
(red line in Fig.~\ref{eeff})
and the angle dependent current through lead $\alpha$ [see (\ref{ieq})]
\begin{eqnarray}
\lefteqn{ I_\alpha(\phi)=}\nonumber\\
&&-\rho_{00}(\Sigma^{\rm in}_{+0,\alpha}+\Sigma^{\rm in}_{-0,\alpha})
+\rho_{++}(\Sigma^{\rm out}_{+0,\alpha}-\Sigma^{\rm in}_{D+,\alpha})\nonumber\\
&&+\rho_{--}(\Sigma^{\rm out}_{-0,\alpha}-\Sigma^{\rm in}_{D-,\alpha})
+\rho_{DD}(\Sigma^{\rm out}_{D+,\alpha}+\Sigma^{\rm out}_{D-,\alpha})\, .
\label{ieq2}
\end{eqnarray}
(red line in Fig.~\ref{iphifig})
Eventually, the Schr\"odinger equation for the torsional degree of freedom with
the effective potential (\ref{veff})  is solved providing eigenfunctions
$\Psi_\nu(\phi)$ and energies $\epsilon_\nu$.
Note that these depend on the electronic steady state of the junction and thus
on bias and gate voltage as well as temperature.
Angle-averaged expectation values of observables $X(\phi)$ are obtained from
$\langle X(\phi)\rangle=\int{\rm d}\phi\sum_\nu\, X(\phi)\,
|\Psi_\nu(\phi)|^2\, \exp(-\beta\epsilon_\nu)/Z$ with partition function $Z$.
A thermal equilibrium state of the {\em torsional} sector is not in conflict
with steady-state solutions of the {\em electronic} sector far from
equilibrium due to the coupling of the former one to a heat bath environment
of residual vibronic modes. Note that this in turn suppresses any possible bistability of the system,
e.g. in Fig.~\ref{eeff} (top) a bistability between the global
minimum situated at $\phi=0$ and the local one around
$\phi=\pi/2$.
With typical energy barriers in $V_{\rm eff}$ of order $eV$ (Fig.~\ref{eeff}),
the mean angle $\langle \phi\rangle$ is dominated over a broad temperature
range mostly by torsional states well localized around the minima of
$V_{\rm eff}$ (green lines in Fig.~\ref{iphifig}).
Finite angle-averaged currents $\langle I(\phi)\rangle$ result
from sufficient overlap of angle dependent currents (\ref{ieq2}) with
these states.

Before we proceed, let us specify the constraints for the above scenario.
It is based on a time scale separation between electronic passage through the
molecule and torsional motion (adiabaticity)  and on a separation between
 electronic transition energies and electrode-molecule couplings
$\hbar\Gamma$ (perturbation theory).  Specifically, this means
$\epsilon_\nu \ll \hbar\Gamma\ll E_+-E_-$ which should be accessible in
experimental set-ups due to $\epsilon_\nu\sim meV\ll E_+-E_-\sim eV$.
\section{Transport properties}
\begin{figure}
\epsfig{file=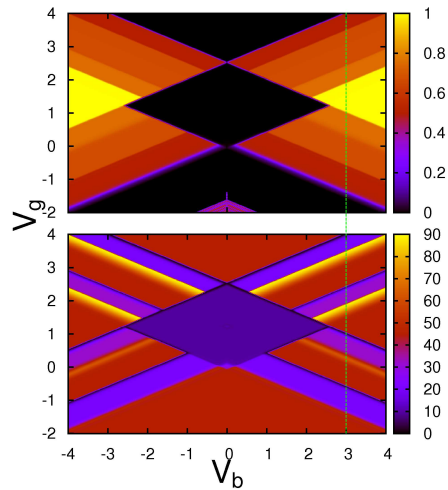, angle=270, width=6.5cm}
\caption{Net current in units of $e\Gamma$ (top) and mean torsional angle
(degrees, bottom)  in  steady state   vs.\ gate voltage $V_g$  and bias voltage
$V_b$ (in units of $V$) with $V_S=0$.}\label{phifig}
\end{figure}

\begin{figure}
\epsfig{file=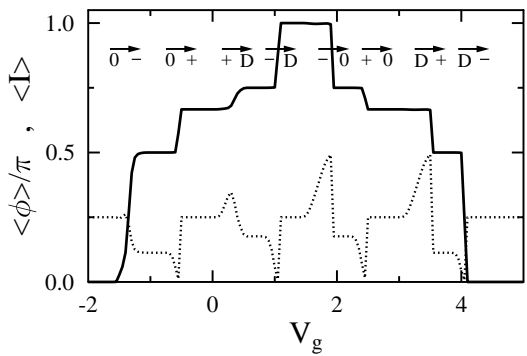, width=7cm}
\caption{Current and equilibrium angle vs.\ $V_g$ at fixed $V_b=3V$,for $\kappa=0$.
Arrows denote the opening (closing) of transition channels.}\label{currentcut}
\end {figure}

 Let us now discuss the steady state properties in
more detail. We start with the case of a vanishing intramolecular field $\kappa=0$.
In Fig.~\ref{phifig} (bottom) the twist angle displays stability plateaus
which reveal conformational changes with increasing bias and gate voltage,
respectively.  In a central diamond (low bias voltage, $V_g>0$) the molecule
resides in an almost planar configuration ($\langle\phi\rangle/\pi\approx 0$),
while outside it is found mainly out of plane with
$\langle\phi\rangle/\pi\approx \frac{1}{4}$ or
$\langle\phi\rangle/\pi\approx \frac{1}{8}$. One additional plateau appears
with $\langle \phi\rangle/\pi \approx \frac{3}{16}$ when four transition
channels are accessible: $|0\rangle\to |\pm\rangle, |\pm\rangle\to |D\rangle$.
This step-wise snapping leaves its imprint on the current as seen in
Figs.~\ref{phifig} (top) and \ref{currentcut}. According to (\ref{ieq2}) and
(\ref{rates}) transport channels open when $eV$ exceeds transition energies
between molecular states which lie in the conduction window determined by $V_g$.
The current exhibits distinct plateaus with $\langle I\rangle/(e\Gamma)=0,
\frac{1}{2}, \frac{2}{3}, \frac{3}{4}, 1$ which can be individually activated
by the gate voltage. An asymmetric response to $V_g$ is also seen with a
central Coulomb diamond shifted towards positive voltages
[\onlinecite{grifoni}].

\begin{figure}
\epsfig{file=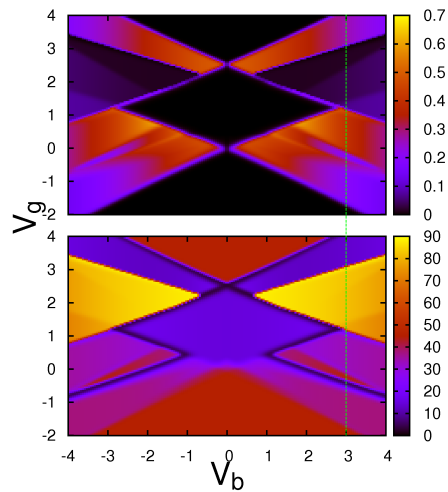, angle=270, width=6.5cm}
\caption{Same as Fig.~\ref{phifig} but with $V_S=0.3V_b$. The vertical line
at $V_b=3V$ marks the cut displayed in Fig.~\ref{cut}.}\label{phistarkfig}
\end{figure}

\begin{figure}
\epsfig{file=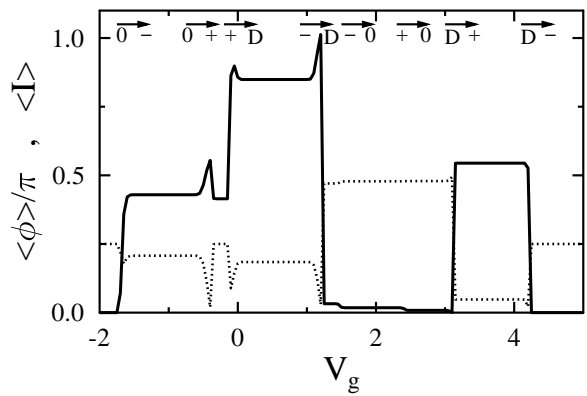, angle=0, width=8cm}
\caption{Current $\langle I\rangle$ in units of $e\Gamma$ (solid) and average
twist angle $\langle \phi\rangle/\pi$ (dashed) at $V_b=3V$ vs.\ $V_g$ (in $V$).
To the left (right) of the maximal current arrows indicate the opening
(closing) of transition channels between molecular states
$|0\rangle, |\pm\rangle, |D\rangle$.}\label{cut}
\end{figure}

In realistic junctions the presence of an intramolecular field (according to
[\onlinecite{xue}] we set $\kappa=0.3$) breaks the symmetry, thus opening a gap between
the $E_+$ and the $E_-$ surfaces (Fig.~\ref{eeff})
and decreasing the HOMO ($E_0$)--LUMO ($E_-$) separation [\onlinecite{xue}].
While for the mean twist angle and the net current (Fig.~\ref{phistarkfig})
the general appearance of the central diamond survives, new wedge-like structures
appear with $\langle\phi\rangle/\pi=1/2$ and thus
 $\langle I\rangle\approx 0$ due to $T(\pi/2)=0$. Namely, for finite Stark shifts
the effective coupling between the leads and the one-particle eigenstates
becomes highly asymmetrical and peaked near $\phi=\pi/2$. This effect
is surprisingly robust and dominates the conduction properties even
at very low internal fields ($\kappa=0.01$). The fully symmetric situation for
$V_S=0$ turns thus out to be extremely unstable and not applicable to actual
junctions.
A cut through the $V_b$-$V_g$ plane at a fixed bias voltage (Fig.~\ref{cut})
reveals the details of the opening and closing of transitions between molecular
states upon sweeping $V_g$ from negative to positive values and the simultaneous
snapping of the dihedral angle.  In contrast, previous treatments where the
torsion angle is considered to be fixed [\onlinecite{venka,xia,mis1,mis2}]
provide conductances $\propto \cos^2(\phi)$. Here, transitions between current
plateaus occur rather abruptly with simultaneous switchings of
the molecular backbone. Notably, this behavior is very insensitive to
temperature fluctuations; the system showed just the same plateaus (albeit
slightly rounded) for $T=300K$.
Of particular interest are sequences $\langle I\rangle/e\Gamma\approx 1
\leftrightarrow 0 \leftrightarrow 1/2$ in the regime $V_g=0\ldots 4$ with
substantial conformational changes
$\langle\phi\rangle/\pi\approx 1/4 \leftrightarrow 1/2\leftrightarrow 0$.
This could be exploited as a molecular switch (current on/off) with the
benefit of being insensitive against variations of temperature and internal
fields in contrast to alternative proposals. In another application
the voltage dependence of the molecular conformation could be used by
operating the biphenyl as a molecular motor [\onlinecite{selden}]. More
specifically, in a first version a slowly varying time periodic gate voltage
at a fixed bias voltage induces a valve-like behavior (e.g.\ open for
$\phi=0$, closed for $\phi=\pi/2$). The state of the rotor is directly read
off from the variations of the dc-current. In an extended set-up side-groups
are attached to each of the biphenyl rings to break in an additional static
electric field the $\pi$-symmetry. A time-dependent gate voltage may then
generate left- resp.\ right-handed rotations of the individual rings. This
situation will be explored in a subsequent work.

\section{Summary}
We have analyzed charge transfer through conjugated organic
molecules contacted to metallic electrodes where a single sluggish vibrational
degree of freedom strongly couples to excess charges. In the regime of higher
voltages and molecule-electrode couplings large compared to vibrational
excitations, but small compared to intramolecular electronic level spacings,
voltage dependent energy surfaces are determined.  Bias and gate voltage allow
to activate current plateaus that correspond to specific molecular
conformations under steady state conditions.
Applications may include non-linear junctions robust against thermal
fluctuations and molecular motors.
\acknowledgments{We thank M. Thoss and C. Timm for valuable discussions.
Financial support from the DFG through SFB569 is gratefully acknowledged.}

\end{document}